# Scattered EXIT Charts for Finite Length LDPC Code Design

Moustafa Ebada, Ahmed Elkelesh, Sebastian Cammerer and Stephan ten Brink
Institute of Telecommunications, Pfaffenwaldring 47, University of Stuttgart, 70569 Stuttgart, Germany
{ebada,elkelesh,cammerer,tenbrink}@inue.uni-stuttgart.de

*Abstract*—We introduce the Scattered Extrinsic Information Transfer (S-EXIT) chart as a tool for optimizing degree profiles of short length Low-Density Parity-Check (LDPC) codes under iterative decoding. As degree profile optimization is typically done in the asymptotic length regime, there is space for further improvement when considering the finite length behavior. We propose to consider the average extrinsic information as a random variable, exploiting its specific distribution properties for guiding code design. We explain, step-by-step, how to generate an S-EXIT chart for short-length LDPC codes. We show that this approach achieves gains in terms of bit error rate (BER) of $0.5\,\text{dB}$ and $0.6\,\text{dB}$ over the additive white Gaussian noise (AWGN) channel for codeword lengths of 128 and 180 bits, respectively, at a target BER of $10^{-4}$ when compared to conventional Extrinsic Information Transfer (EXIT) chart-based optimization. Also, a performance gain for the Binary Erasure Channel (BEC) for a block (i.e., codeword) length of 180 bits is shown.

## I. Introduction

The design of finite length codes has recently received a lot of attention [1][2] due to applications requiring short block lengths such as upcoming communication standards (e.g., the upcoming 5th generation mobile standard proposed by the 3GPP standardization group) as well as sensor networks, i.e., machine-to-machine type communication, and Internet of Things (IoT) networks. Although widely used in many communication standards, Low-Density Parity-Check (LDPC) codes for very short lengths can not be straightforwardly designed with existing tools, as, e.g., the conventional Extrinsic Information Transfer (EXIT) chart-based designs are based on the asymptotic (infinite length) code behavior. Thus, those optimized degree distributions may not be the best possible choice for short length codes. In [3][4], the convergence behavior of finite length turbo codes was analyzed using an "*EXIT band chart*", addressing the infinite length limitation of conventional EXIT charts. For finite length LDPC codes, the decoding trajectories are not perfectly coinciding with the analytical EXIT curves as the assumption of independent messages does not hold. The trajectories are "scattered" and show a quasi-chaotic behavior, as, for example, observed in [5].

However, the *averaged* convergence behavior of short length LDPC codes can be predicted with EXIT charts sufficiently well, as the decoding trajectories tend to be distributed around the analytical EXIT curves (i.e., the analytical EXIT curves describe an average behavior of the decoding trajectories of the different decoding passes).

In this paper, we propose to use "Scattered" Extrinsic Information Transfer (S-EXIT) charts as a guideline to design finite length LDPC codes, i.e., to optimize the degree profile for short length codes. Besides many other useful tools for distance spectrum optimization such as, e.g., the progressive edge growth (PEG)-algorithm [6], the S-EXIT provides a systematic approach to LDPC code degree profile fine-tuning. Before introducing S-EXIT charts, we briefly revisit conventional EXIT chart design.

## II. EXIT Charts for LDPC Codes

EXIT charts were introduced in [7] to study the iterative decoding behavior of two or more constituent decoders in a concatenated code environment. It can be used to predict whether the overall decoding process of an infinite length code will converge or not, and can give an estimate of the lowest signal-to-noise-ratio (SNR) needed for convergence. Additionally, EXIT charts can give an indication about the required number of decoding iterations for convergence, if successful decoding is possible [7]. Furthermore, a bound on the expected maximum a posteriori (MAP) performance as shown in [8] can be derived based on EXIT charts.

A decoding trajectory in the EXIT chart describes the exchange of extrinsic information between the two constituent decoders, where a decoding trajectory tracks the actual simulated information transfer within the free-running iterative decoder. The averaged decoding trajectory typically iterates within the bounded region between the analytical EXIT curves of the two constituent decoders, verifying the derived analytical EXIT curve expressions (i.e., prediction holds) [7][9].

For the inner decoder, the a priori information is initially given by the mutual information of the channel. Next, one inner decoding iteration is performed and the extrinsic (mutual) information from the inner decoder is interleaved and passed on to the outer decoder to become its a priori information. One outer decoding iteration is performed and the extrinsic information is now considered to be the a priori information of the inner decoder, and so on, as can be seen in Fig. 1. Thus, an EXIT curve is an input/output relation of individual constituent decoders. Any EXIT curve represents the output extrinsic information $I_E$ of a constituent decoder after one decoding iteration as a function of the input a priori information $I_A$ [7].

This work has been supported by DFG, Germany, under grant BR 3205/5-1.

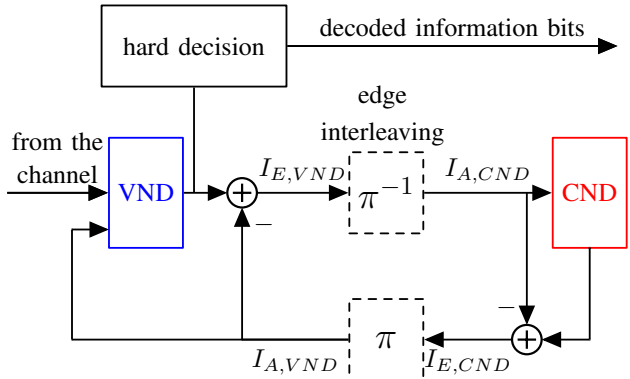

Fig. 1: LDPC iterative belief propagation (BP) decoder.

Any intersection between the EXIT curves of the two decoders in the EXIT chart means that the decoding process will get stuck. Successful decoding is achieved when the trajectory manages to "sneak" through the bottleneck (between the two EXIT curves) and to reach the point $(1,1)$, converging towards low BER (i.e., there exists an open tunnel between the EXIT curves) [7].

Infinite length concatenated codes tailored to iterative decoding can be designed efficiently with the aid of the EXIT chart. The code is optimized by performing a curve fitting operation between the EXIT curves of the two decoders. In other words, the two EXIT curves are matched together such that the *open* tunnel between the two EXIT curves is as small as possible to operate at lowest achievable channel mutual information, i.e., close to channel capacity [7]. For more details, we refer the interested reader to [7], [9].

In this paper, we mainly focus on the design of short length LDPC codes on which our proposed S-EXIT charts are defined and illustrated. However, extending the algorithm to accommodate other types of codes under iterative decoding appears to be possible, and may gain new interesting insights.

Good infinite length LDPC codes can be designed by choosing suitable check node and variable node degree distributions such that the Variable Node decoder (VND) and the Check Node decoder (CND) EXIT curves in the EXIT chart are matched together. This task can now be efficiently solved by using a linear programing approach. However, bit error rate (BER) curves of finite length LDPC codes have a non-abrupt (i.e., more *"smooth"*) waterfall which makes the design based on optimizing the code threshold inefficient, as this method ignores the error floor behavior of the finite length code. It was shown in [10] that a balance between the threshold and error floor performance can be obtained through constraining the optimization problem.

### III. SCATTERED EXIT CHARTS

The S-EXIT chart is a simulation-based tool that uses the statistics of numerous EXIT trajectories obtained from simulations of the actual iterative decoder, and tracks their frequency of occurrence in a two-dimensional histogram over the EXIT mutual information plane [11]. The S-EXIT chart thus enables us to "see through the clutter", in spite of the non-deterministic behavior of the individual EXIT trajectories. It enables to get an insightful expectation of the iterative decoding behavior of the short code via applying some statistical analysis on several instances of it. This results in design guidelines that can be used to further optimize the degree profile of, for instance, short LDPC codes. One can consider the S-EXIT chart as a complementary tool to the classical EXIT chart, useful for short codes as it provides more information about the short code behavior and, thus, helps further optimize its degree distribution.

Note that, this process is also *complementary* to the PEG algorithm [6] and other parity-check matrix optimization algorithms leading to tanner graphs with large girth, as we optimize the degree profile and average over an arbitrarily large set of different randomly constructed H-matrix realizations, so that the effect of its specific realization becomes irrelevant. In other words, the performance shown is an average improvement over arbitrarily large sets of H-matrix ensembles. This can still be combined with other methods to obtain the best instance of a specific code.

*A. Notation*

Due to the finite length of the considered code, different realizations of a channel with a fixed channel condition (i.e., fixed $\epsilon$ for BEC or fixed $\frac{E_b}{N_0}$ for AWGN channel) will exhibit different EXIT behavior. This variation is inversely related to the block length, as will be pointed out later. Unlike the classical EXIT chart where $I_{E,CND}$ (or $I_{E,VND}$) can be represented by a deterministic function of $I_{A,CND}$ (or $I_{A,VND}$), this is not possible for the S-EXIT chart.

$I_{E,CND}$ (or $I_{E,VND}$) can be now regarded as a random variable, with a certain distribution that can be tracked for a given $I_{A,CND}$ (or $I_{A,VND}$) value during the decoding process. A major difference between the work presented here and, e.g., the work in [4] is the consideration of the actual (i.e., simulated) decoding trajectories of the free-running component decoders. Also, in this work, statistical analysis over the obtained results for different channel realizations is applied in the sense that for every single $I_{A,CND}$ (or $I_{A,VND}$), a probability distribution for the sequence of output $I_{E,CND}$ (or $I_{E,VND}$) is obtained. Typically, the mutual information computation is based on histograms. However, for short length codes the histogram suffers from insufficient statistics, causing high variations. For the rest of this work, (1) is used to compute the mutual information $I(\mathbf{L}; \mathbf{X})$ based on an observation LLR vector $\mathbf{L} = \{L_n\}$ and transmitted codeword $\mathbf{X} = \{x_n\}$, where $1 \leq n \leq N$ and $N$ is the code length [12], which holds for any arbitrary LLR-value distribution:

$$\begin{aligned} I(\mathbf{L}; \mathbf{X}) &= 1 - \mathrm{E}\left\{\log_2\left(1 + e^{-\mathbf{L}}\right)\right\} \\ &\approx 1 - \frac{1}{N}\sum_{n=1}^{N} \log_2\left(1 + e^{-x_n \cdot L_n}\right) \end{aligned} \quad (1)$$

*B. Obtaining the S-EXIT chart for a specific code*

The procedure for obtaining an S-EXIT chart for a given LDPC code is, step-by-step, illustrated in Fig. 2. Each step is discussed in detail in the following.

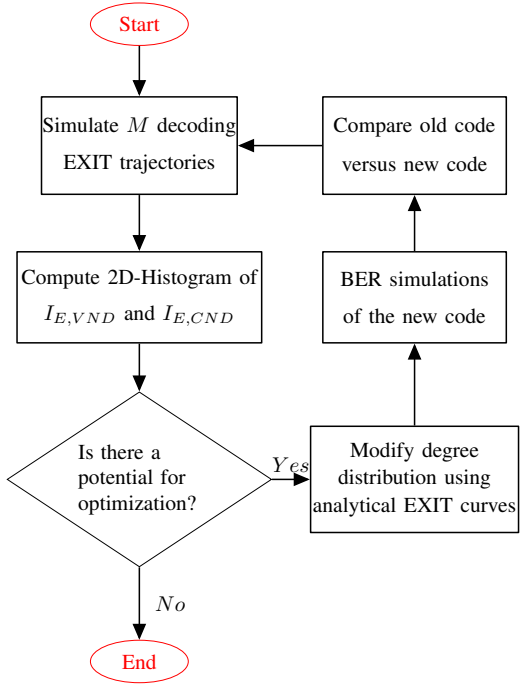

Fig. 2: S-EXIT chart-based degree distribution optimization.

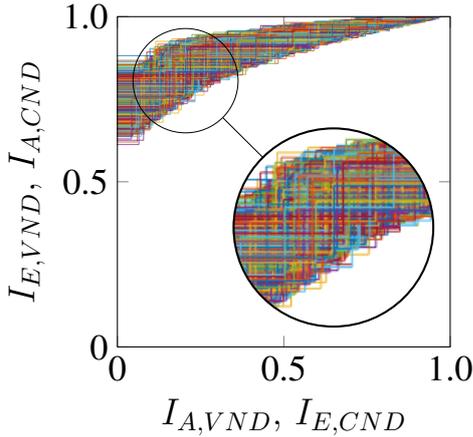

Fig. 3: Acquiring $M$ decoding trajectories of a given code.

*1) Acquiring M simulated EXIT trajectories:* The first phase of generating an S-EXIT chart is to simulate $M$ individual EXIT trajectories of the actual iterative decoder (as shown in Fig. 3), where the realization of the channel is the only variable that changes from one simulation to another (i.e., the codeword might also change, however the BP algorithm is symmetric and, thus, transparent to the chosen codeword). Contrary to classical EXIT trajectories as in Fig. 3, *only the vertices* of the $M$ trajectories are plotted, as depicted in Fig. 4. It is observed that the sample set of points contains two sets: a set of red points representing the CND output, and a set of blue points representing the VND output.

*2) S-EXIT charts as 2D-histograms:* Further statistical analysis is required to acquire knowledge about the distribution of the CND and VND points in the scattered EXIT chart. The simulated distributions result in a new simulation-based chart, as shown in Fig. 7a, 8a and 9a akin to the classical EXIT chart, where they both hold information about the individual CND/VND component decoders. More precisely, the S-EXIT chart is considered as a *weighted* EXIT chart that takes into consideration the frequency of occurrences of the extrinsic information $I_E$ for each specific $I_A$.

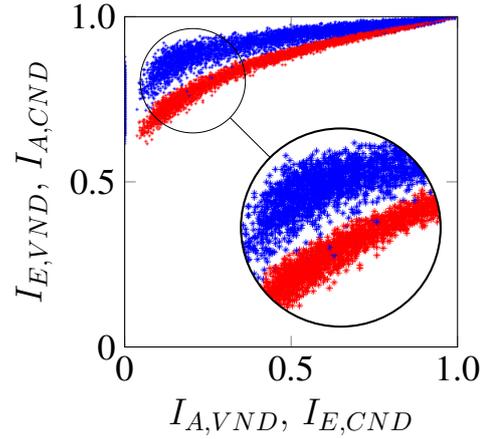

Fig. 4: $M$ scattered decoding trajectories; blue: VND curve, red: CND curve.

*3) Check for optimization potential:* For short codes, the S-EXIT chart *freezes* the randomness in the decoding trajectories of the free-running decoder (i.e., the statistical analysis *eliminates* the effect of the outlier trajectories, and allows the points of the most likely trajectories to dominate).

One then may need to balance reducing the gap between the scattered points of the VND and that of the CND component decoders in specific regions, while reducing the overlap in other regions, thus the code rate is kept constant, in order to reduce the probability of the decoding getting stuck, as in Fig. 7a.

*4) S-EXIT chart guided degree distribution optimization:* Next, the analytical EXIT chart is used as a guideline to modify the degree profile(s) according to one of the aforementioned scenarios, as no analytical description of the S-EXIT chart is available. The major constraint is the rate loss. In other words, it can be formulated as an optimization problem under the constraint that the code rate $R$ is to be kept constant, by, in turn, keeping the area between the two EXIT curves constant [9].

*5) Verifying the results:* The verification step is conducted through two steps:

(i) Obtaining the BER curve for the modified code and comparing with the original code (i.e., before adjusting any of its degree profiles), as in Fig. 7c, 8c and 9e. The modified code may have better BER performance in the low BER region, accompanied by a slightly degraded BER performance in the higher BER region (as for code A in Fig. 7c and the $2^{nd}$ modified version of code C in Fig. 9e), and vice versa, or an enhanced performance over

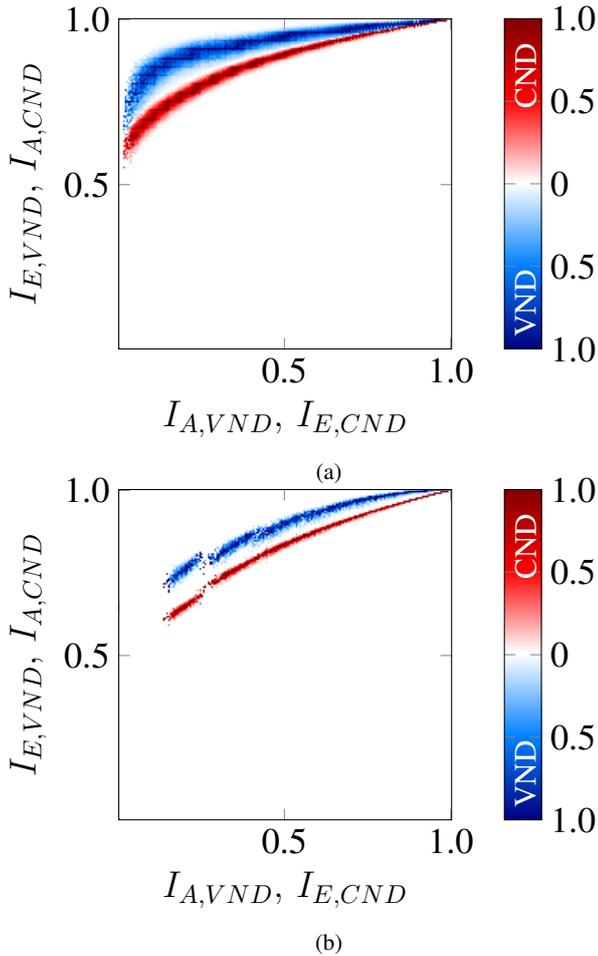

Fig. 5: S-EXIT histogram chart for: (a) $N = 155$, (b) $N = 3000$; parameters: BEC, $\epsilon = 0.25$, $(3,5)$-reg. LDPC code.

- the overall SNR range (as for code B in Fig. 8c and the $1^{st}$ modified version of code C in Fig. 9e).
- (ii) Conducting the same steps from 1 to 3 on the modified code to check for a further optimization potential, see code C optimization process (Fig. 9).

### C. Properties of S-EXIT charts

*1) Effect of codeword length:* The S-EXIT chart shows more variations for shorter length codes than longer ones. Fig. 5 illustrates this effect for a short and a long LDPC code with the given parameters, respectively.

*2) BER region variations:* It is observed that the high BER regions have more variation (i.e., more scattering) than the low BER regions, as seen in Fig. 7a, 8a, 9a and 9c. This observation agrees with the results in [5], where it was shown that the channel dispersion is larger at the beginning of the iterative decoding (i.e., high BER regions) and decreases towards the end of the decoding (i.e., low BER regions).

*3) Dependent vs independent component decoders:* Due to the short length of the code, the assumption that the successive decoding iterations are independent does not hold any more. This is another reason for the wider variations in the high BER regions compared to that of the low BER regions, where the decoding trajectory managed to converge initially. Fig. 6 depicts the S-EXIT chart of a 180 bit $(3,6)$-regular LDPC code over a BEC with $\epsilon = 0.25$, where each single component decoder is individually simulated with an independent input a priori information $I_A$ per simulation point. In other words, Fig. 6 shows the S-EXIT histogram chart of the above-mentioned code where the interdependence between the component decoders is eliminated.

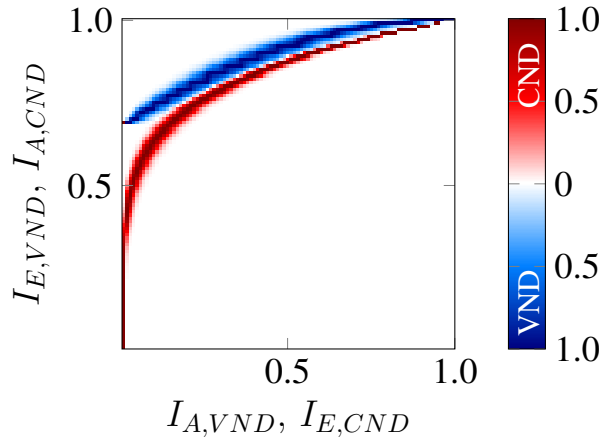

Fig. 6: S-EXIT histogram of a $N = 180$ $(3,6)$-regular LDPC code over a BEC with $\epsilon = 0.25$; each component decoder curve has been simulated independently.

## IV. SIMULATION RESULTS

In this section, the S-EXIT chart is applied to fine-tune three different LDPC codes of short lengths. These codes were originally designed based on the most commonly used design method of matching the VND and CND EXIT curves of the classical EXIT chart, in the same sense as in [13]. The original and modified parameters of the three codes are depicted in Tab. I, Tab. II and Tab. III. Although no comprehensive PEG algorithm has been employed, basic optimizations such as removing girth-4 cycles has been applied to all codes in this paper. The procedure given in section III-B will be applied to the three codes referred to as *code A*, *code B* and *code C* to obtain the modified codes, and for their comparison. The codes are optimized over different channels as indicated in Tab. I, Tab. II and Tab. III. It is worth mentioning that, although the concept is exemplified for the (binary) AWGN channel, any extensions towards other channel front-ends, such as, e.g., quadrature amplitude modulation (QAM) with iterative demapping, are straightforward.

### A. Code A [N=180, BEC]

The parameters of *code A* are shown in Tab. I. This code was optimized for a BEC with $\epsilon = 0.25$. The modified code is obtained after reducing the average variable node degree $\bar{d}_v$ by increasing the weights of the low density nodes, while increasing the average check node degree $\bar{d}_c$ such that the rate is kept constant. The performance of both codes are compared and depicted in Fig. 7. A coding gain is observed when comparing the slightly modified code with the original one.

For the remainder, we focus on optimizing codes over AWGN channels. All S-EXIT charts in this section are plotted using a grid size $N_{grid} = 200$.

TABLE I: Original vs modified degree profiles of *Code A*: BEC, $N = 180$, $R = 0.5$.

| | |
|---|---|
| $\lambda_{orig}(Z)$ | $0.33\,Z^1 + 0.16\,Z^2 + 0.01\,Z^3 + 0.16\,Z^4 + 0.06\,Z^5$ $+ 0.02\,Z^9 + 0.26\,Z^{14}$ |
| $\lambda_{mod}(Z)$ | $0.3\,Z^1 + 0.15\,Z^2 + 0.2\,Z^3 + 0.25\,Z^4 + 0.1\,Z^{14}$ |
| $\rho_{orig}(Z)$ | $0.9\,Z^6 + 0.1\,Z^7$ |
| $\rho_{mod}(Z)$ | $0.01\,Z^1 + 0.02\,Z^2 + 0.1\,Z^3 + + 0.435\,Z^6 + 0.435\,Z^7$ |

*B. Code B [N=128, AWGN]*

The parameters of *code B* are shown in Tab. II. This code was optimized for an AWGN channel at $\frac{E_b}{N_0} = 2\,\mathrm{dB}$. The check and variable node degrees are both kept constant. Only the weight distribution of the variable nodes is a bit altered as depicted in Tab. II. This results in an approximate coding gain of $0.5\,\mathrm{dB}$ at BER of $10^{-4}$.

TABLE II: Original vs modified degree profiles of *Code B*: AWGN, $N = 128$, $R = 0.5$.

| | |
|---|---|
| $\lambda_{orig}(Z)$ | $0.245\,Z^1 + 0.195\,Z^2 + 0.07\,Z^3 + 0.1\,Z^4 + 0.39\,Z^{14}$ |
| $\lambda_{mod}(Z)$ | $0.17\,Z^1 + 0.27\,Z^2 + 0.08\,Z^3 + 0.14\,Z^4 + 0.04\,Z^9 + 0.1\,Z^{11} + 0.2\,Z^{14}$ |
| $\rho_{orig}(Z)$ | $Z^7$ |
| $\rho_{mod}(Z)$ | $Z^7$ |

*C. Code C [N=180, AWGN]*

The parameters of *code C* are shown in Tab. III. This code was optimized for an AWGN channel at $\frac{E_b}{N_0} = 2\,\mathrm{dB}$. The $1^{st}$ modified code is obtained by slightly altering the weight distribution of the variable nodes, similar to *code B*, resulting in a net coding gain of $0.2\,\mathrm{dB}$ at BER of $10^{-4}$. Also, the S-EXIT chart of this new code is plotted, see Fig. 9c, and further optimized by altering both variable and check node average distribution degree, resulting in a net coding gain of $0.6\,\mathrm{dB}$ at BER of $10^{-4}$, when compared to the original *code C*. However, due to the *pronounced* change in the degree distribution, especially in the high BER region as clearly observed in Fig. 9d, a slight performance loss is introduced by this $2^{nd}$ modified code in the SNR range between $1\,\mathrm{dB}$ and $2.75\,\mathrm{dB}$, see Fig. 9e.

TABLE III: Original vs modified degree profiles of *Code C*: AWGN, $N = 180$, $R = 0.5$.

| | |
|---|---|
| $\lambda_{orig}(Z)$ | $0.245\,Z^1 + 0.195\,Z^2 + 0.07\,Z^3 + 0.1\,Z^4 + 0.39\,Z^{14}$ |
| $\lambda_{1st}(Z)$ | $0.205\,Z^1 + 0.345\ Z^2 + 0.016\,Z^3 + 0.082\,Z^8 + 0.008\,Z^9 + 0.344\,Z^{14}$ |
| $\lambda_{2nd}(Z)$ | $0.125\,Z^1 + 0.25\,Z^2 + 0.25\,Z^3 + 0.375\,Z^{10}$ |
| $\rho_{orig}(Z)$ | $Z^7$ |
| $\rho_{1st}(Z)$ | $Z^7$ |
| $\rho_{2nd}(Z)$ | $0.75\,Z^7 + 0.25\,Z^8$ |

## V. Outlook and Conclusion

We introduced the concept of S-EXIT charts as a tool for short length code design. The idea extends "classic" EXIT charts such that the variance of mutual information is taken into account during the degree profile optimization. An improvement in terms of BER up to $0.5\,\mathrm{dB}$ and $0.6\,\mathrm{dB}$ over the AWGN channel for block lengths of 128 and 180 bits was achieved, respectively, at a target BER of $10^{-4}$ when compared to the conventional EXIT chart-based results. Also, a performance gain for LDPC code design over the BEC for a block length of 180 bits has been demonstrated. For future work, the S-EXIT optimized degree profiles can be combined with advanced code (i.e., graph) constructions such as the PEG algorithm taking into account the girth optimization of the code. Another extension of this work may be using S-EXIT charts to obtain an estimation of the BER of a finite length code based on the simulated S-EXIT chart, and not only bounds on the error probability. Also empirical VND and CND curves (i.e., worst case scenarios) can be inferred from the points in the S-EXIT chart, and degree profile optimization can be adjusted accordingly.

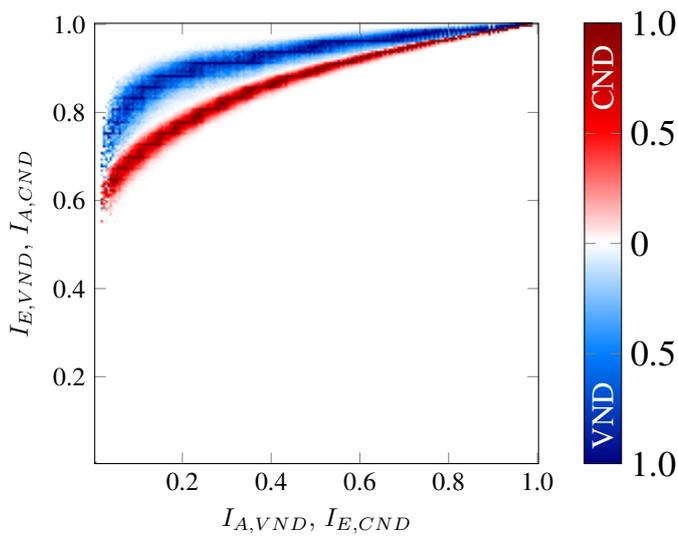

(a) S-EXIT histogram of the original code.

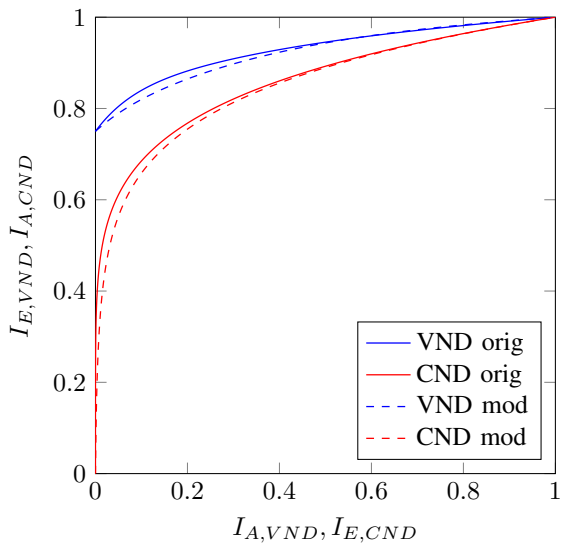

(b) Analytical EXIT curves based on the degree profiles of the original and the modified code.

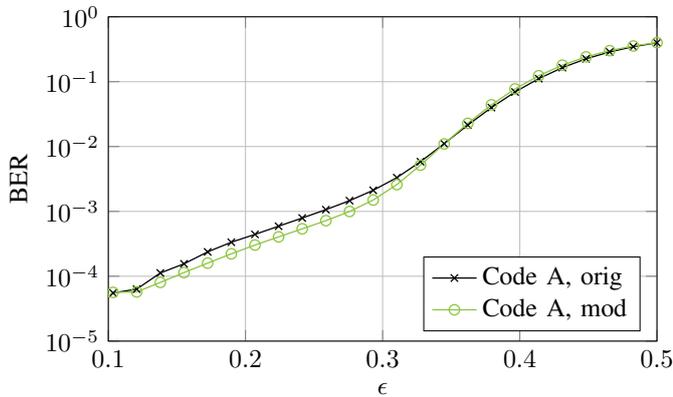

(c) BER comparison averaged over arbitrarily large set of different randomly constructed ensembles.

Fig. 7: Code A over BEC.

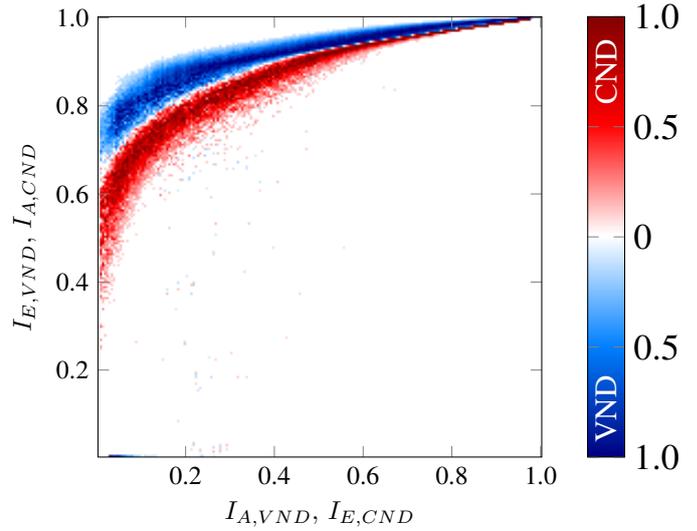

(a) S-EXIT histogram of the original code.

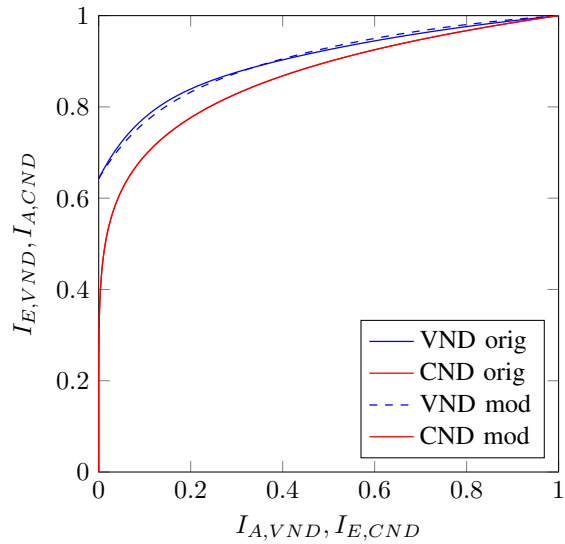

(b) Analytical EXIT curves based on the degree profiles of the original and the modified code.

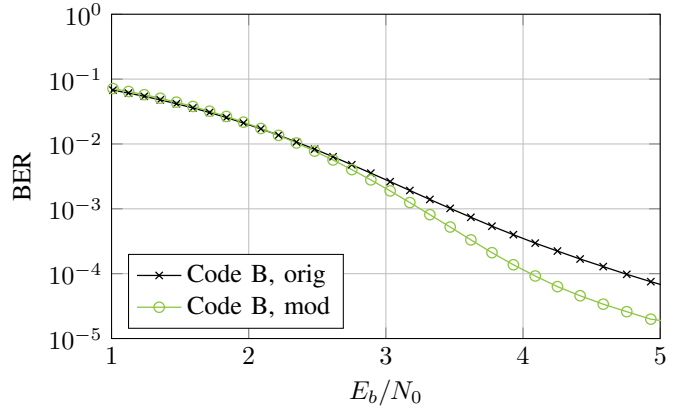

(c) BER comparison averaged over arbitrarily large set of different randomly constructed ensembles.

Fig. 8: Code B over AWGN.

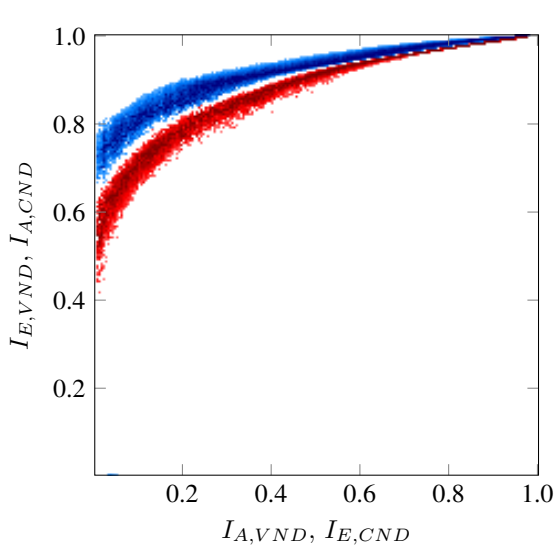

(a) S-EXIT histogram of the original code.

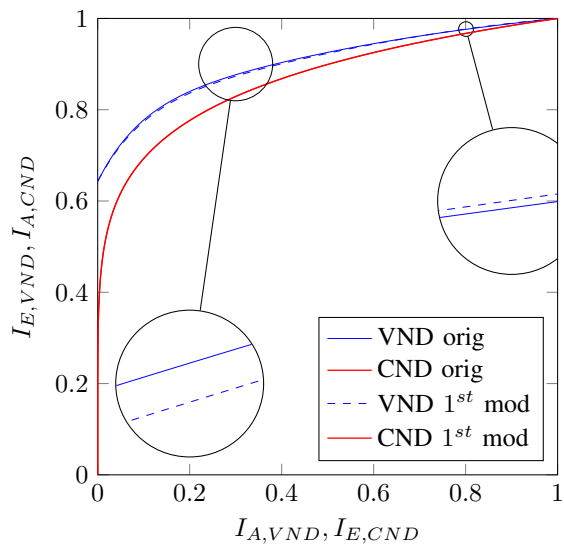

(b) Analytical EXIT curves based on the degree profiles of the original and the $1^{st}$ modified code.

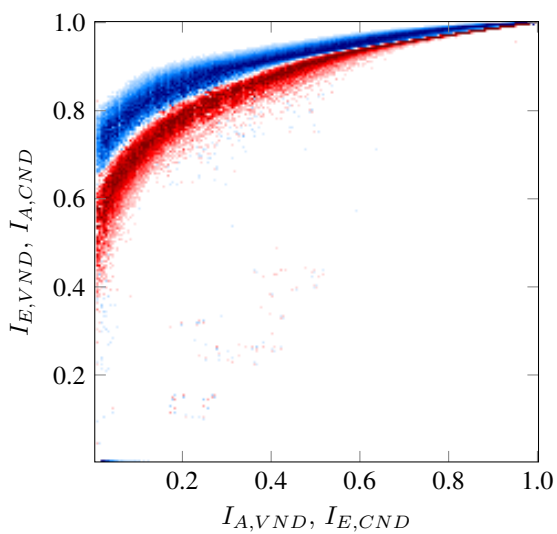

(c) S-EXIT chart of $1^{st}$ modified code.

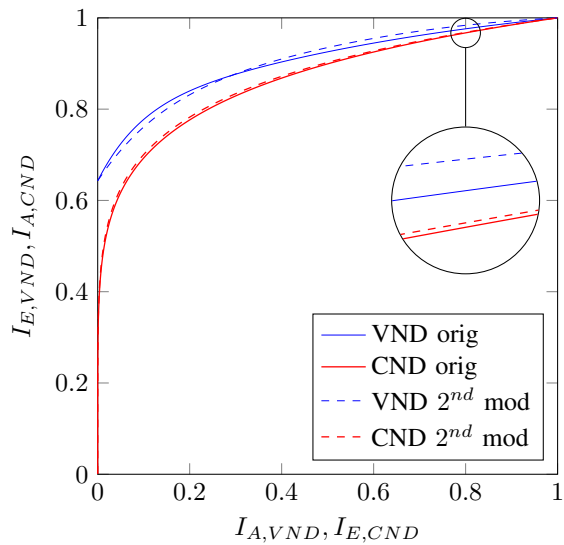

(d) Analytical EXIT curves based on the degree profiles of the original and $2^{nd}$ modified code.

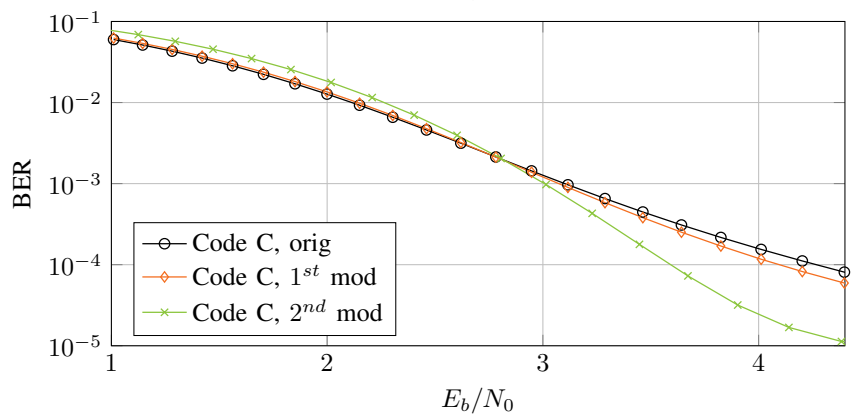

(e) BER comparison averaged over arbitrarily large set of different randomly constructed ensembles.

Fig. 9: Code C over AWGN.